\documentclass[]{article}

\usepackage{epstopdf}
\usepackage{subfigure}
\usepackage{graphicx}
\usepackage[numbers,sort&compress]{natbib}
\bibpunct[, ]{[}{]}{,}{n}{,}{,}

\begin{document}



\title{Effect of stacking fault energy on nucleation limited plasticity in Cu-Al alloys}

\author{G. Kamalakshi, Prachi Limaye, \\
M.P. Gururajan~and Prita Pant~\thanks{$^\ast$Corresponding author. Email: pritapant@iitb.ac.in}\\ \\
{\em Department of Metallurgical Engineering and Materials Science,}\\
{\em Indian Institute of Technology Bombay, Mumbai, India}}

\maketitle

\begin{abstract}
We study the effect of Stacking Fault Energy (SFE) on the deformation behaviour of copper  and copper-aluminium  alloys using Molecular Dynamics (MD) simulation. We find that both yield stress and the magnitude of stress drop at yield decrease with increasing Al content. This “anomalous” softening behaviour is explained on the basis of nucleation controlled yielding behaviour. Further, the decrease in stress drop is rationalised in terms of the stored energy available at yielding – we show that this decreases with increasing Al. As a result, the maximum dislocation density  is found to decrease with increasing Al content.
\end{abstract}


\section{Introduction}

We are interested in studying the effect of Stacking Fault Energy (SFE) on deformation of single crystal fcc materials. Several 
Molecular Dynamics (MD) studies have been carried out to investigate the effect of SFE
on plastic deformation in fcc metals~\cite{DiffSFECu,DiffSFEAl,Nw3fcc,ncmetals} and 
alloys~\cite{Goldalloy,Alalloy,TJRupert,Nisuperalloy,CuAg,RKRajgarhia}.
Of the studies on  alloys, gold alloys are studied by 
nano-indentation~\cite{Goldalloy}, the studies on Ni-base superalloys are with specific reference to 
dislocation-solute interactions~\cite{Nisuperalloy}; the studies on alloyed aluminium~\cite{Alalloy}, Cu-Ag~\cite{CuAg} 
and Cu-Pb~\cite{TJRupert} systems are on nanomaterials. Rajgharia et. al~\cite{RKRajgarhia} have studied the effect 
of solute addition on tensile deformation of Cu-Sb single crystals; however, the amount of solute added 
and the corresponding change in SFE are very small (upto 2 at\% and about 1 mJ/m$^2$, respectively). 
Our interest is to study the effect of a large 
change of SFE with alloying addition on tensile deformation in fcc single crystals (albeit simulated using 
periodic boundary conditions).  

We have chosen Cu-rich Cu-Al alloys (specifically, Cu-4.59at.\%Al, Cu-8.94at.\%Al and 
Cu-13.07at.\%Al alloys) as the model system for our MD studies 
because the alloy remains as a single phase fcc upto about 13 at. \%Al~\cite{phasedia} 
and the SFE varies from about 43 mJ/m$^2$ for 
pure copper to about 6 mJ/m$^2$ for nearly 13 at.\% Al alloy~\cite{Rohatgi,sfelit1,sfelit3,sfelit2,CarterRay}. 

We find that in the nucleation limited systems, decreasing SFE with increasing solute content the onset of 
plastic deformation takes place at lower and lower stresses and the
magnitude of the drop in stress at yield keeps decreasing. We show that this is due to the increasing ease 
of nucleation of loops of partial dislocations with decreasing SFE. 

\section{Simulation details}

We have used LAMMPS~\cite{lammps} to carry out the MD simulations (of tensile deformation) using the Embedded Atom 
Method (EAM) potentials of Zhou and Ward~\cite{ZW}. All simulations are carried out at 300 K and 0 Pa (using an NPT ensemble); 
the temperature and pressure are maintained using Nos\'{e}-Hoover thermo- and baro-stats respectively. The
temperature is ramped from 100 K to 300 K for the simulations. 

We have used a simulation box of $30\times30\times30$ unit cells (108000 atoms) along x, y and z axes with 
periodic boundary conditions in all three directions; the x, y and z axes of the simulation cell are oriented 
along [100], [010] and  [001] crystallographic directions respectively
and the loading is done along the z-axis. After several trials, we have found that a strain rate of $10^{8}$ $s^{-1}$
is optimal since it leads to the required strains after about 15 million time steps and does
not result in unphysical fluctuations in the stress-strain plot. The reported values of elastic moduli,
yield stress, and drop at yield point, are based on an average of three simulations.

We have calculated the components of the stiffness tensor (that is, $C_{11}$, $C_{12}$ and  $C_{44}$ 
since we are working with a cubic system) and lattice parameters 
for pure Cu and the Cu-Al alloys at 300 K. The elastic constants are calculated using
the  {\em change box} command of LAMMPS as indicated in the solved example~\cite{EConstant}, 
and the values are given in Table.~\ref{CandY} along with the lattice parameters.

The microstructures and dislocation structures are analysed using Ovito~\cite{ovito}.

\begin{table}[bthp]
\caption{Lattice parameter (a) and the elastic constants ($C_{11}$, $C_{12}$ and  $C_{44}$)  of Cu and Cu-Al alloys.}
\begin{center}
{\begin{tabular}[l]{@{}lcccccc}
\hline \hline
System &   $a$ ($A^\circ$)& $C_{11}$ (GPa)  & $C_{12}$ (GPa)&   $C_{44}$ (GPa)   \\
\hline \hline
  \small  Pure Cu            & \small   3.612$\pm0.0$  & \small 201.34$\pm0.00$  & \small  126.11$\pm0.00$ & \small  79.52 $\pm0.00$       \\ 
  \small Cu-4.59at.$\%$Al & \small  3.626$\pm0.0$ & \small 160.83$\pm0.40$  & \small  121.86$\pm0.07$ & \small  69.59$\pm0.25$        \\ 
  \small Cu-8.94at.$\%$Al & \small  3.639$\pm0.0$ & \small 153.23$\pm0.20$  & \small  119.93$\pm0.09$ & \small 64.49$\pm0.14$        \\
  \small Cu-13.07at.$\%$Al& \small  3.652$\pm0.0$ & \small  145.87$\pm0.47$ & \small  117.02$\pm0.52$& \small   59.83$\pm0.82$       \\
 \hline \hline 
 \end{tabular}}
 \end{center}
 \label{CandY}
\end{table}

\section{Results and discussion}

\begin{figure}
\begin{center}
\subfigure[]{
\resizebox*{5cm}{!}{\includegraphics{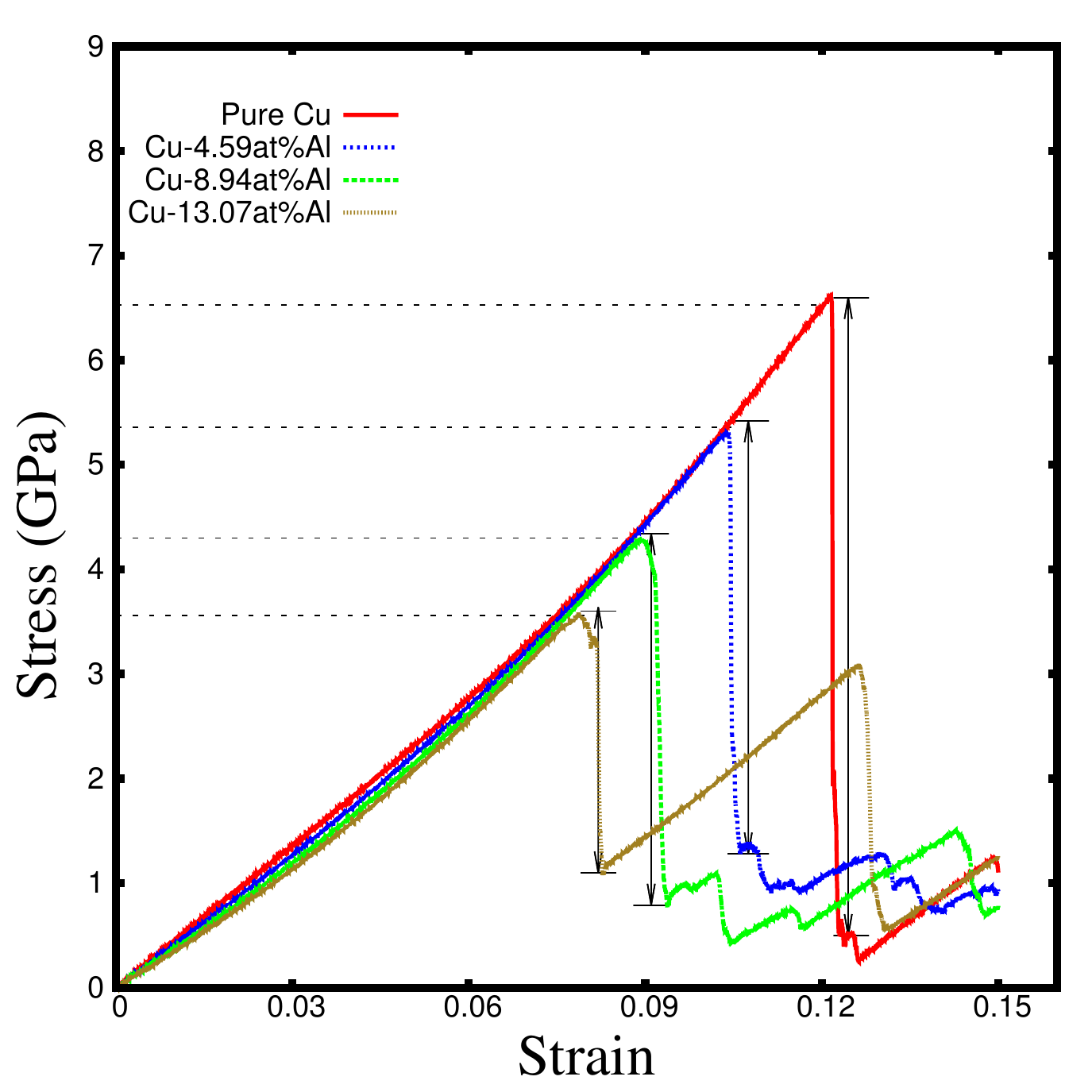}}}\hspace{5pt}
\subfigure[]{
\resizebox*{5cm}{!}{\includegraphics{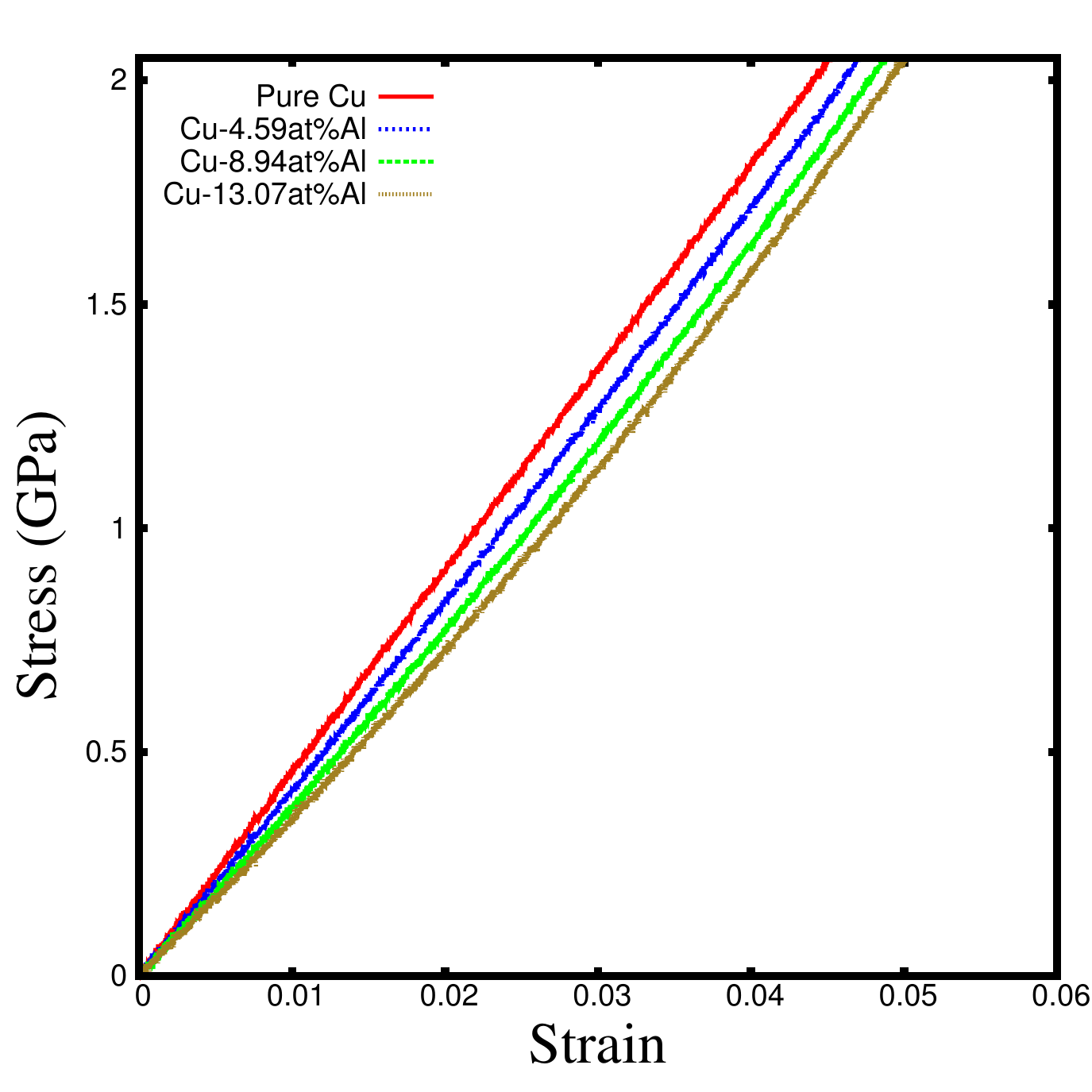}}}
\caption{(a) Stress-strain response of Cu and Cu-Al alloys, black dotted lines show the yield point and black arrows show the amount of drop in stress at yield point. (b) Linear portion of the stress strain plot.}\label{stress-strainZWard}
\end{center}
\end{figure}

In Fig.~\ref{stress-strainZWard}, we show the stress-strain response for pure Cu, and Cu-4.59at.\%Al, Cu-8.94at.\%Al and 
Cu-13.07at.\%Al alloys. In the inset, we also show the initial portion of the stress-strain plots from which it can be seen
that the Young's modulus ($E$) decreases with increasing Al concentration; that is, the alloys are becoming
less stiff with increasing Al. In the figure, in all plots, we have marked the first abrupt drop in stress by black dotted
lines. At this point, as we show below, the onset of platicity by nucleation of dislocation loops is seen. Hence, we call this
stress as the yield stress, $\sigma_y$.  The yield stress values decrease with increasing Al; further, the amount of drop in stress 
(indicated by black arrows) also decreases with increasing Al. In Table.~\ref{YSandDropTable}, we have listed these yield 
stresses and the magnitude of drop in stress at these points.
In other words, the addition of aluminium makes the material not only compliant but also soft. 

 \begin{table} 
 \caption{Variation of yield stress, drop in stress ($\Delta \sigma$) and Youngs modulus ($E$) for Cu and Cu-Al alloys}
 \begin{center}
 {\begin{tabular}[l]{@{}lcccccc|} 
 \hline \hline
 Material &  $\sigma_y$ (GPa)  &   $\Delta{\sigma}$ (GPa) & E (GPa)    \\
 \hline \hline
  Cu            & \small  6.58$\pm0.02$       & \small  6.08$\pm0.02$ & \small 53.07$\pm 0.16 $   \\
  Cu-4.59at.$\%$Al & \small  5.29$\pm0.04$       & \small  4.22$\pm0.18$  & \small 50.69$\pm  0.24$   \\
  Cu-8.94at.$\%$Al & \small  4.13$\pm0.16$       & \small  3.29$\pm0.20$   &\small 47.83$\pm  0.30$  \\
  Cu-13.07at.$\%$Al& \small  3.52$\pm0.05$       & \small  2.45$\pm0.60$   & \small  45.85$\pm 1.59 $   \\
\hline \hline
 \end{tabular}}
 \end{center}
 \label{YSandDropTable}
\end{table}

\subsection{Reasons for decreasing $\Delta \sigma$}

\begin{figure}[htpb]
\centering
\includegraphics[trim=0.5cm 0.5cm 0.5cm 0.5cm,height=3in,width=3in]{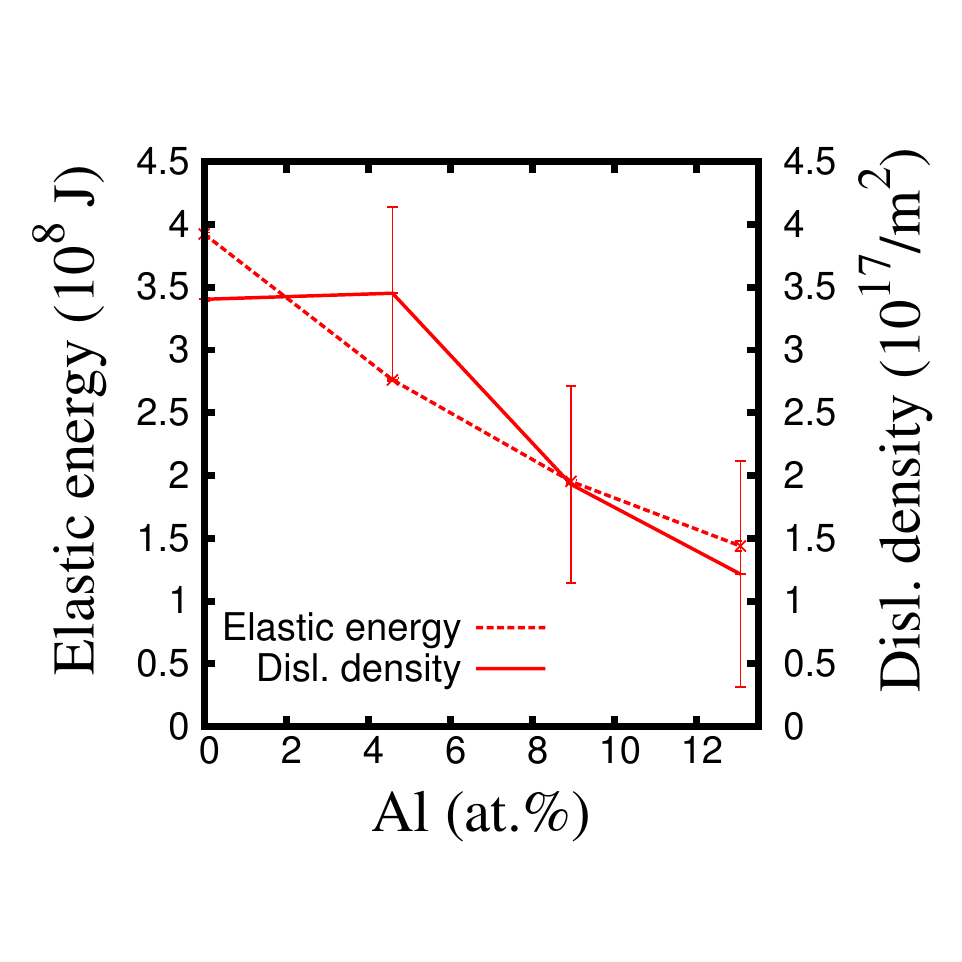} 
\caption{Variation of elastic energy and total dislocation density with Al.}\label{density1}
\end{figure}

It is easier to explain as to why $\Delta \sigma$ decreases with increasing
Al. As the Al at.\% increases, the alloys become more compliant; in addition, as we show below, it is also easier to 
initiate plastic deformation in them because of the reduced SFE; hence, they store relatively less elastic
energy. We have calculated the slope of the stress-strain curves as the Youngs modulus ($E$). Based
on these moduli, we have calculated the elastic energy as $\sigma_y^2/2E$ where
$\sigma_y$ is the yield stress. As shown in
Fig.~\ref{density1}, the stored elastic energy at yield decreases with increasing Al. So, at the yield point, 
with increasing Al, the system has less and less energy to release in the form of dislocations; and, thus, the 
dislocation densities are expected to decrease with increasing Al content. In Fig.~\ref{density1}, we have also
plotted the dislocation density after the first drop and this is indeed the case. 

\begin{figure}
\begin{center}
\subfigure[]{
\resizebox*{5cm}{!}{\includegraphics{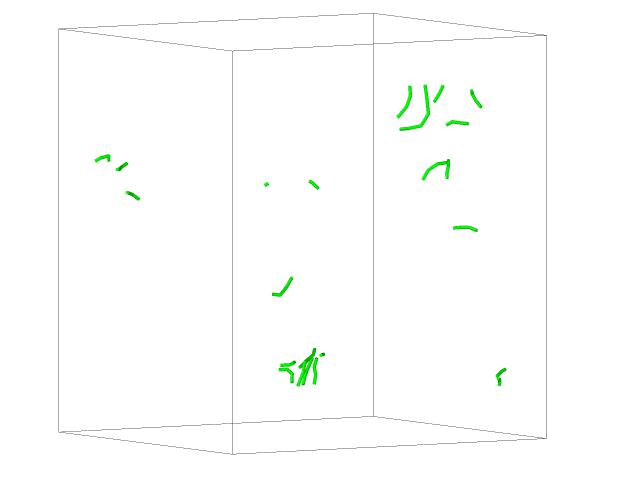}}}\hspace{5pt}
\subfigure[]{
\resizebox*{5cm}{!}{\includegraphics{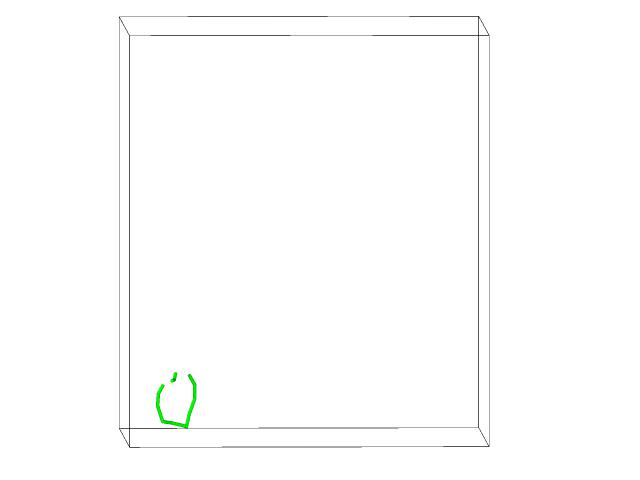}}}
\subfigure[]{
\resizebox*{5cm}{!}{\includegraphics{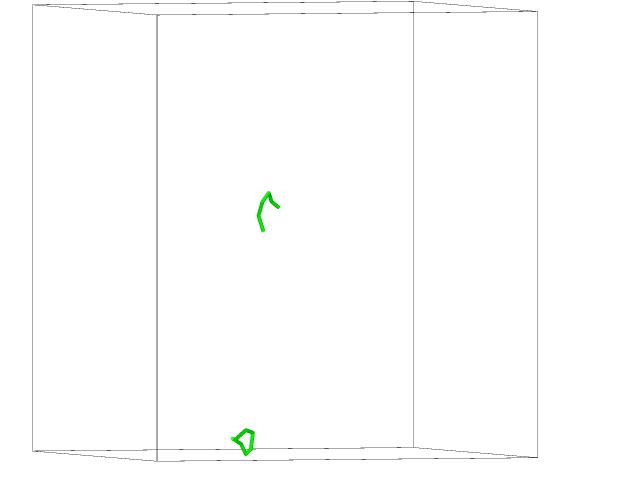}}}\hspace{5pt}
\subfigure[]{
\resizebox*{5cm}{!}{\includegraphics{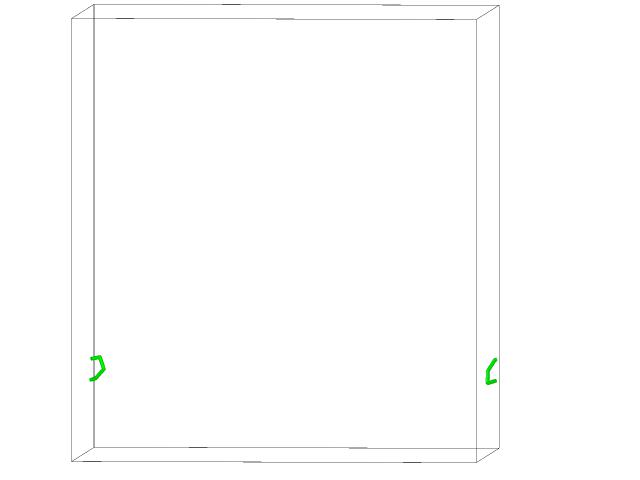}}}
\caption{Nucleation of Shockley partial loop in (a) pure Cu,  (b) Cu-4.59 at.\% Al, (c) Cu-8.94 at.\%Al and
(d) Cu-13.07at.\%Al. The ratios of the length  of Shockley partial dislocations ($L_S$)  to the  length of total dislocations ($L_T$)  near the drop for Cu, Cu-4.59 at.\% Al, Cu-8.94 at.\%Al and Cu-13.07at.\%Al  are 0.967$\pm0.024$, 0.997$\pm0.005$, 1.0$\pm0.0$ and  0.991$\pm0.01$ respectively. }\label{LoopFormation}
\end{center}
\end{figure}

\subsection{Reasons for decreasing $\sigma_y$}

Conventionally, the addition of a solute is expected to increase the yield stress in alloys~\cite{Dieter,Courtney,MayersChawla}.
In our case, we see an anomalous behaviour -- namely, ``solid solution softening". We show below that this ``softening" is
due to the ease of formation of partial loops with increasing Al - because it leads to decreasing SFE.

Figures.~\ref{density1}.a-d, show the dislocation structure at the yield point for Cu and Cu-Al alloys. As is clear from
these figures, in all cases, Schockley partial loops are formed. As a matter of fact, near the yield point,  the ratio of the 
length  of Shockley partial dislocations ($L_S$)  to total dislocation length ($L_T$) is found to be 0.95.

Given the predominance of Schockley partial dislocations in our MD simulations, we have used the continuum model
 proposed by Aubry et al~\cite{Sylvie} to estimate the stress required for the homogeneous nucleation of dislocation loops.
In this model, the energy required for homogeneous nucleation of a partial dislocation loop ($E_{nuc}$) is given as follows:
\begin{equation}\label{disl-nucleation}
 E_{nuc}(\tau) = -b \tau A + \gamma_{SFE} A + 2 \pi R \frac{\mu b^{2}}{8 \pi} \left\lbrace \frac{2-\nu}{1-\nu}\left[\ln\left({\frac{8 R}{r}}\right)-2\right]
 +0.5\right\rbrace,
\end{equation}
where $\mu$  is the shear modulus ($C_{44}$), $R$ the radius of the dislocation loop, $A$ the area of the dislocation loop, 
$b$ the Burgers vector of the Shockley partial ($\frac{1}{6}[11\bar{2}]$), $\nu$ the Poisson's ratio,  $r$ the 
dislocation core radius, $\gamma_{SFE}$ the stacking fault energy and  $\tau$ the applied shear stress. 
In Eq.~\ref{disl-nucleation}, the first term is the elastic energy dissipated by the nucleation of a 
dislocation loop; the second term is the energy of the stacking fault created by the (partial) dislocation loop; and the
third term is the elastic energy of the loop. The energy barrier for the dislocation nucleation is zero when the 
energy dissipated is equal to the sum of the energy of the (partial) dislocation loop and the stacking fault created
by the loop. Therefore, the shear stress required for 
dislocation nucleation $\tau_{nuc}$is obtained by equating Eq.~\ref{disl-nucleation} to zero:
 
\begin{equation}\label{FinalEq1}
  \tau_{nuc} = \frac{\mu b}{4 \pi R} \left\lbrace \frac{2-\nu}{1-\nu}\left[\ln\left({\frac{8 R}{r}}\right)-2\right]
 +0.5\right\rbrace+\frac{\gamma_{SFE}}{b}
\end{equation}
Note that in deriving the above expression, we have replaced $A$ by $\pi R^2$.
Replacing $\mu$  by $C_{44}$ and  $\nu$ by $\frac{C_{12}}{C_{11}+C_{12}}$,  Eq.~\ref{FinalEq1} becomes:
\begin{equation}\label{FinalEq}
  \tau_{nuc} =  \frac{C_{44} b}{4 \pi R} \left\lbrace \frac{2C_{11}+C_{12}}{C_{11}}\left[\ln\left({\frac{8 R}{r}}\right)-2\right]
 +0.5\right\rbrace+\frac{\gamma_{SFE}}{b}.
\end{equation}

To evaluate the above expression, we have used the $\gamma_{SFE}$ values reported in the literature and listed in 
Table.~\ref{bv1}. The values of the elastic constants $C_{11}$, $C_{12}$ and $C_{44}$ and the Burgers vector (through
the lattice parameters) are calculated using MD simulations and are listed in Table.~\ref{CandY}. We have used 
the dislocation core radius ($r$) and radius of the dislocation loop ($R$) as fitting parameters. The $\tau_{nuc}$
values are calculated for (111) slip plane and $\frac{1}{6}[11\bar{2}]$ Burgers vector. 

\begin{figure}[htpb]
\centering
\includegraphics[trim=0.5cm 0.5cm 0.5cm 0.5cm,height=2.5in,width=2.5in]{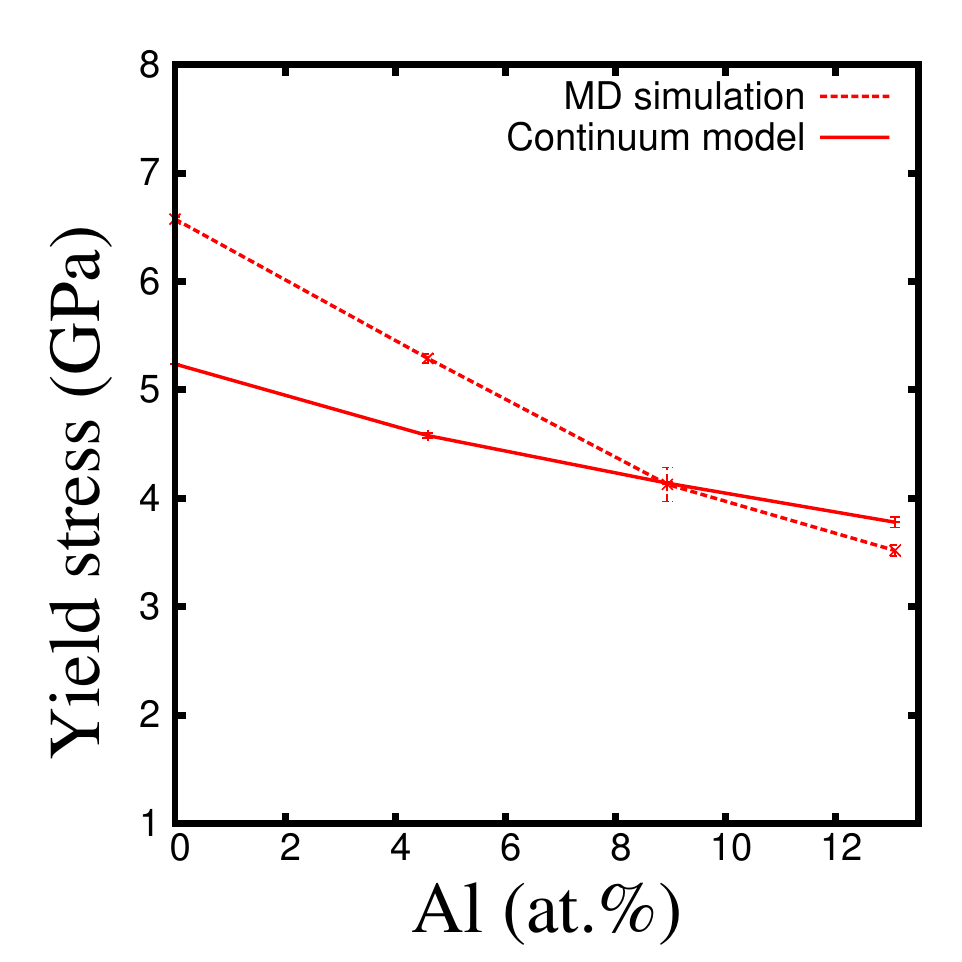} 
\caption{Change in yield stress with Al addition, calculated using continuum model~\cite{Sylvie} and obtained from MD simulations. }\label{YS_both}
\end{figure}

The tensile stress corresponding to $\tau_{nuc}$ was calculated for [001] loading direction using 
$\sigma_{nuc} =  \tau_{nuc}\cos{\phi}cos{\lambda}$ (where $\phi$ and $\lambda$ are the angles
made by the slip plane normal and the Burgers vector, respectively with the loading direction)
and is given in column 4 of Table.~\ref{bv1}. 
It is clear from the Table.~\ref{bv1} that, the tensile stress required for nucleation of partial 
dislocation decreases with increasing Al. In Table.~\ref{bv1}, we also show
the yield stress obtained from the MD simulations which agree fairly well with continuum calculations using
the fitting parameters of $r= 3b = 0.442$ nm 
and $R= 2$ nm.  The variation of yield stress with Al  
from the calculations and MD simulations is shown Fig.~\ref{YS_both}.

\begin{table}[htbp]
 \caption{Stress required for nucleation of partial dislocation with Burgers vector $\frac{1}{6}[11\bar{2}]$ in (111) plane
 for loading along $\langle 001 \rangle$ direction; $\sigma_{nuc}$ is calculated from the continuum model and $\sigma_y$
 is obtained from MD simulations. }
 \begin{center}
{\begin{tabular}[l]{@{}lcccccc}
\hline \hline
 Material& $\gamma_{SFE }$ & $\tau_{nuc}$ &    $ \sigma_{nuc}$   &   $\sigma_y$  \\ 
\hline \hline
  Cu          & \small   43~\cite{CarterRay}  & \small  2.47$\pm 0.0  $ & \small 5.24$\pm 0.0  $ & \small 6.58$\pm0.016$  \\                
  Cu-4.59$\%$Al & \small   25~\cite{Rohatgi} & \small  2.16$\pm 0.01  $   & \small 4.58$\pm 0.02  $      & \small      5.29$\pm0.04$                    \\                                                         
  Cu-8.94$\%$Al& \small   13~\cite{Rohatgi}& \small  1.95$\pm 0.0  $    & \small 4.14$\pm 0.01  $   & \small       4.13$\pm0.16$                     \\                                                         
  Cu-13.07$\%$Al& \small    6~\cite{Rohatgi}  & \small  1.78$\pm 0.02 $   & \small 3.78$\pm  0.05 $      & \small       3.52$\pm0.05$                 \\                                                        
\hline \hline
 \end{tabular}}
 \end{center}
 \label{bv1}
\end{table}

\begin{figure}[htbp]
\begin{center}
\subfigure[]{
\resizebox*{5cm}{!}{\includegraphics{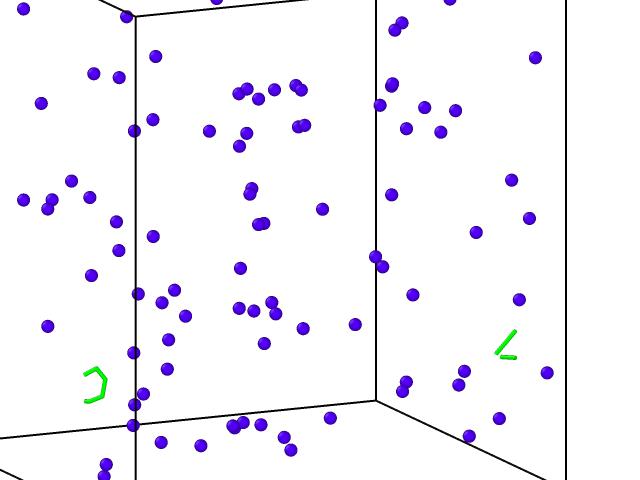}}}\hspace{5pt}
\subfigure[]{
\resizebox*{5cm}{!}{\includegraphics{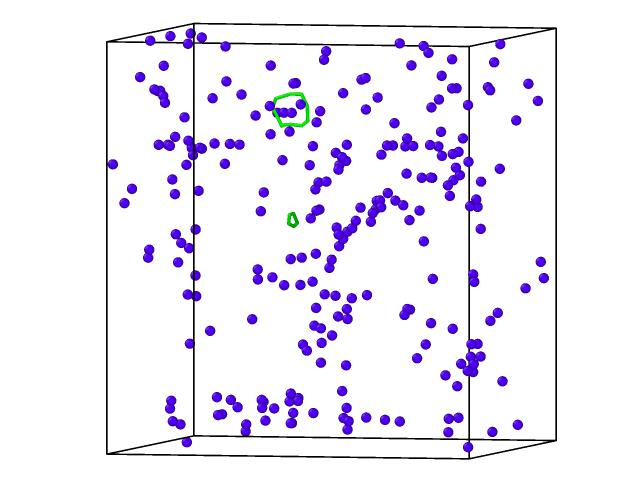}}}
\end{center}
\caption{Nucleation of Shockley partials (shown in green -- using dislocation analysis (DXA) of Ovito) 
in (a) Cu-4.59 at.\% and (b) Cu-8.94 at.\%Al systems. The blue circles are the Al atoms; for the sake of 
clarity, copper atoms are not shown. As is clear from the figure, both homogeneous and heterogeneous 
nucleation are seen. }\label{Alcluster}
\end{figure}

Solid solution softening has been reported in MD simulations of Cu-Pb system by  Rupert~\cite{TJRupert}, Cu-Ag system
by Amigo et al~\cite{CuAg}, and Cu-Sb system by 
Rajgarhia et al~\cite{RKRajgarhia}. Rupert reports that in nanocrystalline materials yield strength is proportional 
to the Young's modulus similar to the metallic glasses, and since the addition of Pb to Cu reduces the Young's modulus, 
the material shows softening behaviour. But our studies are on single crystals where no sources of dislocations
such as grain boundaries exist; and so his explanation is not 
relevant for the system under study. Amigo et al and Rajgarhia et al~\cite{CuAg,RKRajgarhia} find that the local 
strains created by the solute atoms lead to the heterogeneous nucleation of dislocations in their vicinity. However,
in our study, we find nucleation both at Al sites and away from Al sites (even in systems with 4.93 and 8.94 at.\% Al
cases) -- see Fig.~\ref{Alcluster}; thus, we see both homogeneous and heterogeneous nucleation of dislocations and 
hence, the softening in our case can not be attributed to heterogeneous nucleation alone. Interestingly, we have found
that in those studies in which the SFE of copper and aluminium was changed by using different interatomic potentials,
as the SFE is decreased, the deformation is dominated by partials~\cite{DiffSFECu,DiffSFEAl}; specifically,
in the case of copper bicrystals, the same trend, namely decrease in yield stress with decreasing SFE is observed.

\section{Conclusion}

The role  of stacking fault energy on the deformation behaviour of Cu rich Cu-Al alloys, where SFE decreases with increasing Al content, is studied using molecular dynamics simulations. Since SFE decreases, it is easier for the system to nucleate partial dislocations; therefore, yield occurs at lower applied stress with increasing Al content. Because of decreasing Youngs modulus 
and yield stress, the system stores less elastic energy and hence the dislocation density decreases with Al addition. The yield stress values obtained from the simulations agree fairly well with the stress required for homogeneous nucleation of partials calculated using a continuum model. 

\subsection{Acknowledgements}

We thank Mr. Saransh Singh and Dr. Arijit Roy for useful discussions; we thank 
Dendrite, Spinode -- the DST-FIST HPC facility, and C-DAC, Pune 
for computational resources and DST, Government of India for funding this project (14DST017). 

\bibliographystyle{tfq}

\bibliography{kamalakshiguruprita}

\begin{thebibliography}{10}
\newcommand{\printfirst}[2]{#1}
\newcommand{\switchargs}[2]{#2#1}
\providecommand{\url}[1]{\normalfont{#1}}
\providecommand{\urlprefix}{Available at }

\bibitem{DiffSFECu}
V. Borovikov, M.I. Mendelev, and A.H. King, \emph{{Effects of stable and
  unstable stacking fault energy on dislocation nucleation in nano-crystalline
  metals}}, Modell. Simul. Mater. Sci. Eng. 24 (2016), pp. 85017--14.

\bibitem{DiffSFEAl}
T. Shimokawa, A. Nakatani, and H. Kitagawa, \emph{{Mechanical Properties
  Depending on Grain Sizes of Face-Centered-Cubic Nanocrystalline Metals Using
  Molecular Dynamics Simulation ∗ ( Investigation of Stacking Fault Energy '
  s Influence )}}, JSME Int. J. 47 (2004), pp. 1708--1715.

\bibitem{Nw3fcc}
H.S. Park, K. Gall, and J.A. Zimmerman, \emph{{Deformation of FCC nanowires by
  twinning and slip}}, J. Mech. Phys. Solids 54 (2006), pp. 1862--1881.

\bibitem{ncmetals}
V. Yamakov, D. Wolf, S.R. Phillpot, A.K. Mukherjee, and H. Gleiter,
  \emph{Deformation-mechanism map for nanocrystalline metals by molecular-
  dynamics simulation}, Nat. Mater. 3 (2004), pp. 43--47.

\bibitem{Goldalloy}
Y. Li, A. Goyal, A. Chernatynskiy, J.S. Jayashankar, M.C. Kautzky, S.B.
  Sinnott, and S.R. Phillpot, \emph{{Nanoindentation of gold and gold alloys by
  molecular dynamics simulation}}, Mater. Sci. Eng. A 651 (2016), pp. 346--357.

\bibitem{Alalloy}
S. Hocker, M. Hummel, P. Binkele, H. Lipp, and S. Schmauder, \emph{{Molecular
  dynamics simulations of tensile tests of Ni-, Cu-, Mg- and Ti-alloyed
  aluminium nanopolycrystals}}, Comp. Mater. Sci. 116 (2015), pp. 32--43.

\bibitem{TJRupert}
T.J. Rupert, \emph{{Solid solution strengthening and softening due to
  collective nanocrystalline deformation physics}}, Scripta Mater. 81 (2014),
  pp. 44--47.

\bibitem{Nisuperalloy}
X. Zhang, H. Deng, S. Xiao, X. Li, and W. Hu, \emph{{Atomistic simulations of
  solid solution strengthening in Ni-based superalloy}}, Comp. Mater. Sci. 68
  (2013), pp. 132--137.

\bibitem{CuAg}
N. Amigo, G. Guti{\'{e}}rrez, and M. Ignat, \emph{{Atomistic simulation of
  single crystal copper nanowires under tensile stress: Influence of silver
  impurities in the emission of dislocations}}, Comp. Mater. Sci. 87 (2014),
  pp. 76--82.

\bibitem{RKRajgarhia}
R.K. Rajgarhia, D.E. Spearot, and A. Saxena, \emph{{Heterogeneous dislocation
  nucleation in single crystal copper–antimony solid-solution alloys}},
  Modell. Simul. Mater. Sci. Eng. 17 (2009), pp. 55001--13.

\bibitem{phasedia}
B. Lu, K. Chen, and W.J. Meng, \emph{{Quantification of Thermal Resistance of
  Transient-Liquid- Phase Bonded Cu / Al / Cu Interfaces for Assembly of
  Cu-Based Microchannel Heat Exchangers}}, J. Micro Nano-Manuf. 1 (2016), pp.
  1--10.

\bibitem{Rohatgi}
A. Rohatgi, K.S. Vecchio, and G.T. Gray, \emph{{The influence of stacking fault
  energy on the mechanical behavior of Cu and Cu-Al alloys: Deformation
  twinning, work hardening, and dynamic recovery}}, Metall. Mater. Trans. A 32
  (2001), pp. 135--145.

\bibitem{sfelit1}
M. Chassagne, M. Legros, and D. Rodney, \emph{{Atomic-scale simulation of screw
  dislocation/coherent twin boundary interaction in Al, Au, Cu and Ni}}, Acta
  Mater.  (2011), pp. 1456--1463.

\bibitem{sfelit3}
Z.X. Wu, Y.W. Zhang, and D.J. Srolovitz, \emph{{Dislocation – twin
  interaction mechanisms for ultrahigh strength and ductility in nanotwinned
  metals}}, Acta Mater. 57 (2009), pp. 4508--4518.

\bibitem{sfelit2}
M.D. Sangid, T. Ezaz, and H. Sehitoglu, \emph{{Energetics of residual
  dislocations associated with slip – twin and slip – GBs interactions}},
  Mater. Sci. Eng. A 542 (2012), pp. 21--30.

\bibitem{CarterRay}
C.B. Carter and I.L.F. Ray, \emph{{On the stacking-fault energies of copper
  alloys}}, Philos. Mag. 35 (1977), pp. 189--200.

\bibitem{lammps}
 {S. Plimpton}, \emph{{Fast Parallel Algorithms for Short-Range Molecular
  Dynamics}}, J. Comput. Phys. 117 (1995), pp. 1--19,
  \urlprefix\url{http://lammps.sandia.gov}.

\bibitem{ZW}
L. Ward, A. Agrawal, K.M. Flores, and W. Windl, \emph{{Rapid Production of
  Accurate Embedded Atom Method Potentials for Metal Alloys}}, Model. Simul.
  Mater. Sc.  (2012).

\bibitem{EConstant}
A. Thompson. \urlprefix\url{http://lammps.sandia.gov/doc/Section example.html}.

\bibitem{ovito}
A. Stukowski, \emph{{Visualization and analysis of atomistic simulation data
  with OVITO – the Open Visualization Tool}}, Modelling Simul. Mater. Sci.
  Eng. 18 (2009), pp. 015012--7, \urlprefix\url{http://ovito.org/}.

\bibitem{Dieter}
G.E. Dieter, \emph{Mechanical Metallurgy}, 3rd ed., McGraw Hill Education
  (India) Private Limited, New Delhi, 2016.

\bibitem{Courtney}
T.H. Courtney, \emph{Mechanical Behavior of Materials}, 2nd ed., McGraw Hill
  Custom Pub., United States of America, 2000.

\bibitem{MayersChawla}
M.A. Meyers and K.K. Chawla, \emph{Mechanical Behavior of Materials}, 2nd ed.,
  Cambridge University Press, New York, 2009.

\bibitem{Sylvie}
S. Aubry, K. Kang, S. Ryu, and W. Cai, \emph{{Energy barrier for homogeneous
  dislocation nucleation: Comparing atomistic and continuum models}}, Scripta
  Mater.  (2011), pp. 1043--1046.

\end{thebibliography}

\end{document}